# HOW STRUCTURE AFFECTS POWER-LAW BEHAVIOR

HAN   JING [a]        LI   WEI [b]

[a]*University of Science and Technology of China*

[b]*Hua-Zhong Normal University, P.R. China*

Complex systems contain a lot of individuals and some interactions between them. The structure of interactions can be modeled to be a network: nodes represent individuals and links to be interaction between two individuals.   This paper tries to investigate how structure of interactions in the system will affect the power-law behaviors based on the BS co-evolution model from points of system size, the density of connectivity and the ways of connectivity of the interaction network. The current experiments show that small size, high density of connectivity and random connected way will weaken the power-law distribution.

## I. Introduction

Systems which exhibit scale-free power-law behaviors are ubiquitous in physics, biology, geology and even sociology, etc. There are many different ways of generating power-law, self-organized criticality is one among them.

Self-organized Criticality (SOC) manifests that open, dynamical, far-from equilibrium systems consisting of many constituents may evolve towards a critical state without any control from outside agents. In the critical state, a small local perturbation may spread to the whole system through domino-like effects and form an avalanche. Sizes, spatial and temporal as well, of avalanches at the critical state obey a stable scale-free power-law.

However, sometimes people will find that the distribution is not exactly power-law: has falling down tail (see Fig.1), combine with exponential distribution in the tail area (see Fig.2), etc.

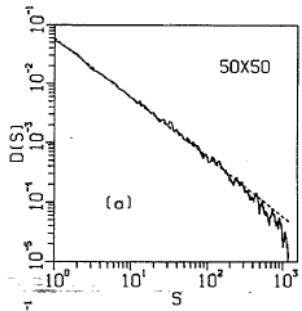   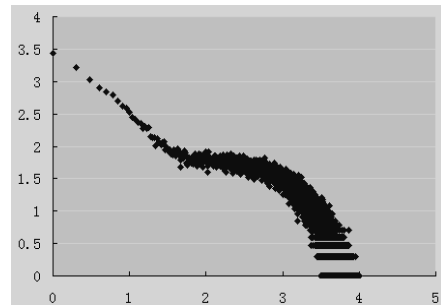

Fig.1 Power-law with falling down tail                   Fig.2. Combined with power-law and Exp. Distribution.

How this happen?    Are they power-laws?

Another example is Bak-Sneppen (BS) evolution model. In this simple model, an ecosystem consisting of many interactive species located on a d-dimensional lattice is studied. Initially, random numbers chosen from a flat distribution between 0 and 1 are assigned to each species as its fitness. At each time step, the species with the smallest fitness (random number) and its 2d nearest neighbors are mutated by being assigned new random numbers taken from the same uniform distribution. This updating process continues indefinitely until the system enters a stable state under which the random numbers in the system are uniformly distributed between a certain threshold and 1. At such a state, a single mutation of a given species may lead to a chain of mutations and thus form an avalanche. Sizes of avalanches follow a simple power-law (Fig.3).

However, in reality, species are not always exactly located in a d-dimensional lattice. Extremely, in an *n* species system, if we change the neighborhood, say, each species has n-1 neighbors, which means the interaction structure is fully connected, size of avalanches will obey an exp. distribution (Fig.4). From this, we know the connected way of the interaction sometimes will affect the power-law behaviors. But how about the middle cases, middle connectivity?

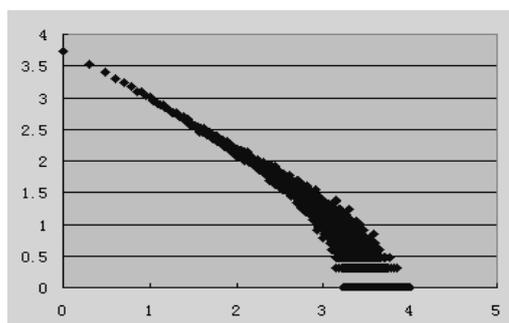
Fig. 3 Power-law in the lattice case

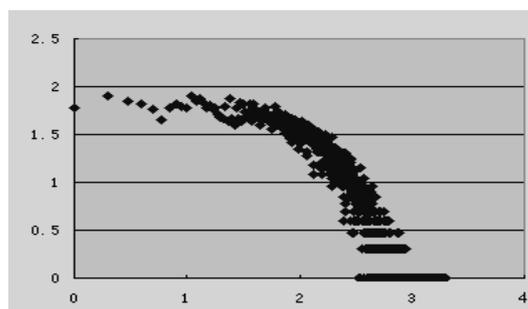
Fig. 4  Exp. in fully connected cases

This paper we try to investigate what affects the power-law behaviors. We propose to make the following investigations:

**What does structure mean in an evolution model?**

**Will structure have some effects on the dynamics of the system, e.g., will there be power-laws of avalanche size distribution?**

## II. What is structure?

Before the investigation, it is necessary to make clear what structure stands for, in an evolution model. To extract the interaction structure from the system, we do the following way: a species (individual) => a node, interaction between two species => link between these two nodes.

We argue that structure may include many factors overlapping to a certain degree. In this article, three main factors that we will take into account are

**System size *n***, number of species in the system,

**Connectivity density *d***, the number of neighbors for each species,

**Types of connectivity *t***, such as *circle*, *lattice*, *hive*, *random*, etc.

The latter two factors represent degrees of local interaction. We will do computer simulations to model evolution model, based on BS model, with various structures.

## III. Experimental Investigations

### 1. System Size

It is obvious that system size will have some effects on the dynamics of the evolution model. Actually, when the system size is finite, there will be finite-size effect. We will roughly divide systems into three categories: small-sized, middle-sized, and large-sized, though it is hard to distinguish them very exactly.

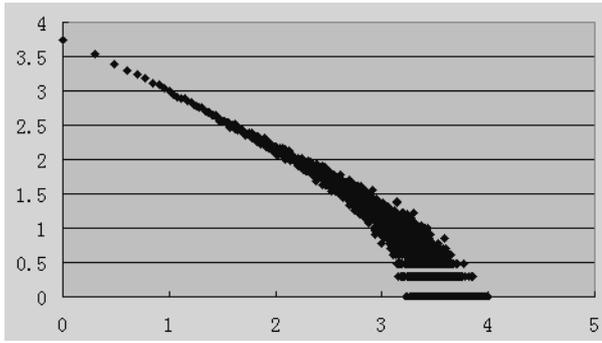

Fig. 5 (size=500, threshold =0.62, 100,000 avalanches)

First, we come to systems of large size. Simulations are done for systems of size 100, 500 and 1000. Avalanches are observed through threshold 0.62. The total number for statistics is 100,000. The simulation result for system of size 500 is presented in Fig. 5. Similar figures can be found for systems of size 100 and 1000. We see that power-law behaviors are explicit in large-sized systems.

Second, we performed simulations on evolution of small-sized systems. The sizes here we chose are 1, 2, and 3, respectively. Actually, sizes 1, 2, and 3 are very special, since in systems of such sizes, all species in the system will mutate at each time step. In some sense, systems of such sizes share the same statistics features, corresponding to the features of flat distribution of random numbers. It is easily derived that sizes of "avalanches" (if can be called, different from those in critical systems) obey exponential distributions. The distributions can be written as,

$$P(S) \sim f_0^{\,S},$$

where $f_0$ is a threshold for defining an avalanche. Theoretical analysis confirms simulation results. Fig. 6 shows "avalanche" size distribution of system of size 3. We can see clearly an exponential distribution.

Third, we come to the evolution of middle-sized systems. Here, we chose sizes 4 and 10. Simulation result for system of size is shown in Fig.4.

It can be seen from Fig. 7 and Fig. 8 that avalanche size distributions of middle-sized systems

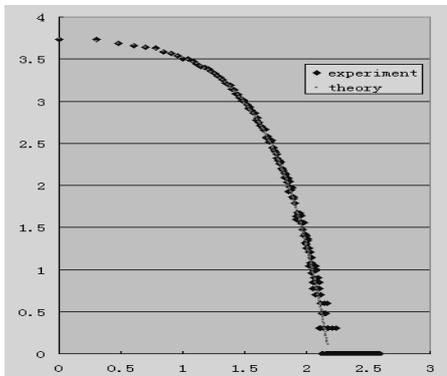

Fig. 6 (size=3, threshold=0.62, 100,000 avalanches)

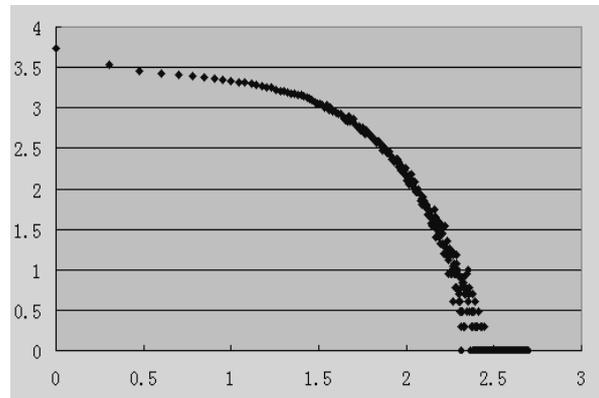

Fig. 7 (size=4, $f_0$=0.62, 100000 avalanches)

combine the features of the exponential distribution and those of power-law distribution. As

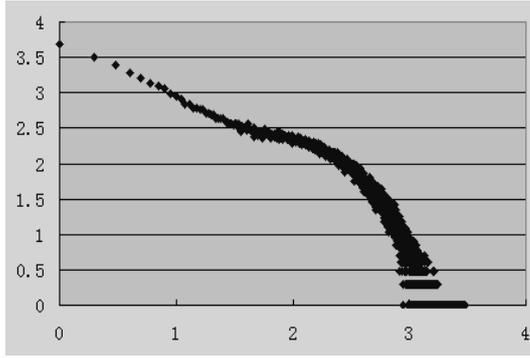 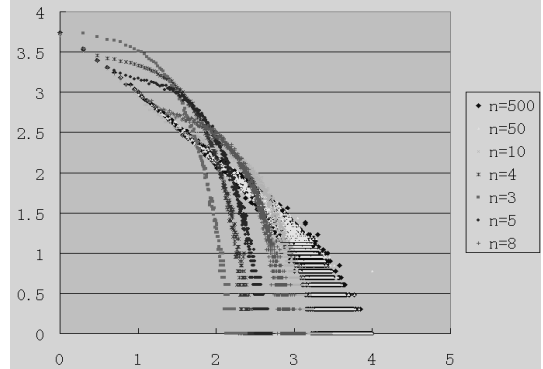

Fig. 8 (size=10, threshold=0.62, 100,000 avalanches)

Fig. 9 Comparison of avalanche size distribution for different sized systems

seen, in the small avalanche size region, sizes of avalanches obey power-law distribution, and in large avalanche size region, they will obey exponential distribution.

The above four figures combined, we can see that as system sizes increase, the avalanche size distribution tends to be power-law more likely. We may conjecture that the distribution transits from exponential ones to power-law ones as we increase the system size. One candidate reason is that increase in system size decreases the effect of random process. This can be seen clearly from the small-sized systems. In such systems, random processes dominate. Hence, for small-sized systems the distribution tends to behave in ways that may bring the systems into the state with maximum entropy.  Comparisons of avalanche size distribution for systems of different sizes are shown in Fig. 9.

## 2. Density of connectivity

In this article, as a simple study, we would like to define density of connectivity as the number with which each individual connects to other individuals in the same system. Hence, for BS model, the density of connectivity is $d=2$, 4 and 6, for 1-, 2-, and 3-dimensional space, respectively. That is to say, for d-dimensional BS model, the density of connectivity is 2d. If for a system of size L, each constituent interacts with all other constituents in the same system, the density of connectivity $d$ is (L-1). We call such a density of connectivity complete connectivity. It can be inferred that as density of connectivity increases, the locality of interaction will be lowered. The interaction in the system will become very complex, and a constituent can affect constituents at a far distance directly. This kind of interaction may not need spreading.

It can be easily inferred that systems with complete connectivity resembles evolves like a random process since at each time step each individual will change simultaneously. After some algebra, the size distribution of avalanches sizes can be written as an exact exponential distribution,

$$P(S) \sim f_0^S,$$

where $f_0$ is a threshold. Simulations of different sized systems with complete connectivity are shown in Fig. 10. These results confirm the theoretical analysis.

We performed simulations on evolution models with density of connectivity being 3. The results are presented in Fig. 7. As seen, the size distribution of avalanches deviates from power-law slightly.  We argue that density of connectivity does have some effects on the forming of power-law behavior.  As density of connectivity increases, the size distribution of avalanches will deviate power-law more likely. Such a conjecture may need more proofs.

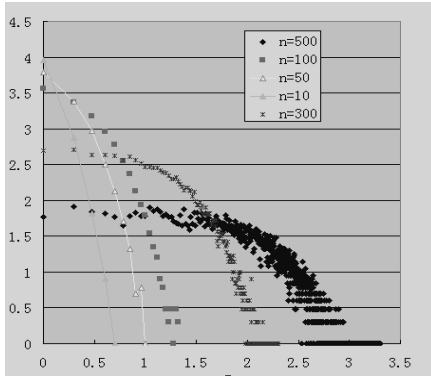 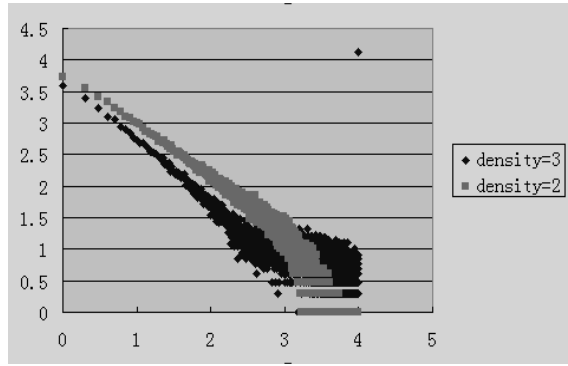

Fig. 10 Complete connectivity     Fig. 11 size=30, threshold=0.62, 100,000 avalanches

A very important question is what the intrinsic mechanism of the deviation from power-law behaviors is. One possible explanation is that the increase in density of connectivity lowers the locality, thus leads to the dominance of randomness. Randomness is crucial in generating exponential distributions.

### 3. Types of connectivity

Even with the same density of the connectivity, there will be different ways to connect. For example, when the density of connectivity is 3, one can add a random link to each node based on the *circle* structure (see Fig.12) to get a network with density of 3, or like a *hive* structure (see Fig.13).

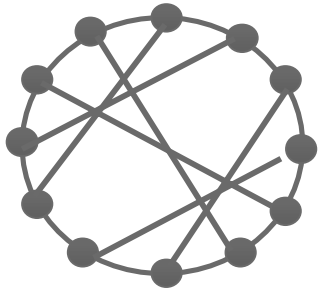 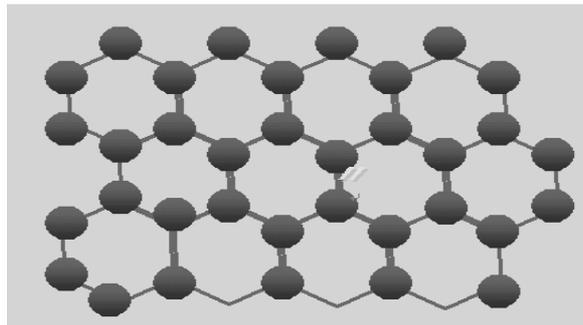

Fig11. Circle + Random          Fig12. Hive

We do some simulation on these two types of structure for system size of 100. Avalanches are observed through threshold 0.40. The total number for statistics is 100,000. We see that power-law behaviors are more explicit in *hive* structure system than the *random* one (Fig. 13).

From this case, we can conclude that the type of the connectivity will affect the power-law behaviors. When we look into the reason of why this happen, we found that the *hive* network is more highly clustered than the *random* one, this also means the interaction in *hive* network spreads slower than that in the *random* network. Does it mean highly clustered networks will has more tendency of power-law behavior than those un-clustered ones?

We consider another example, for *d*=4, two types: 2-dimemsional *lattice* and r*andom* based on *circle*. The computer simulation result is show in Fig. We can also see that the avalanches size-frequence distribution of 2-dimemsional *lattice,* which is a highly clustered network, looks more like power-law than that of the *random* structure, un-clustered network.

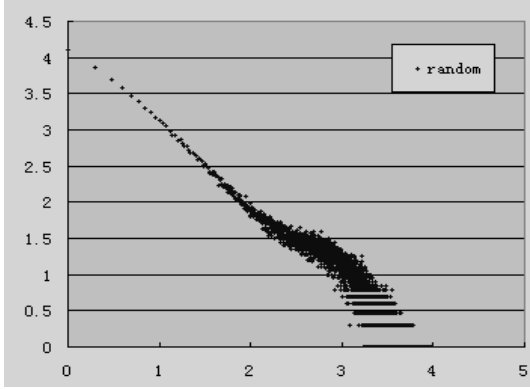
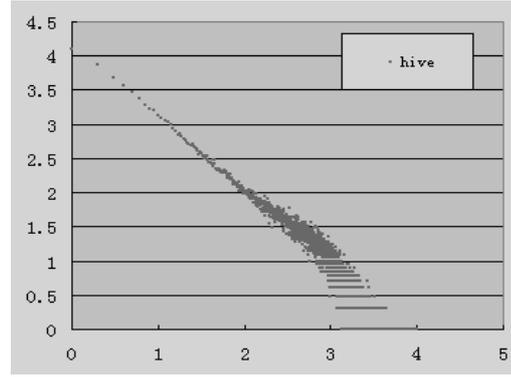

Fig 13 (a) Circle + Random     Fig 13.(b) Hive structure

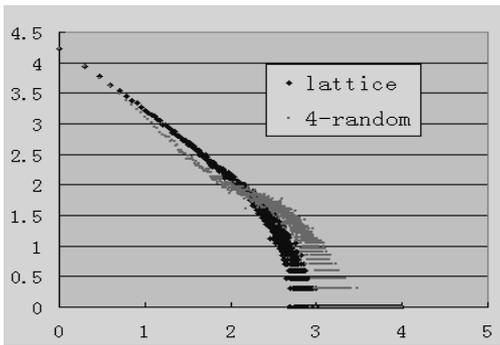

Fig.14 $n$=100, $f_0$=0.30, 10000 avalanches

It seems that highly clustered or not has great effect on the power-law behavior. However, when we transfer a highly clustered network to a highly un-clustered network according to the *Small World* theory, e.g., randomly select a few links from a *hive* or *lattice* network and change them to be un-clustered network, we get confused results. In the hive case, after transfer to a "small world", the size-frequency distribution of avalanches obviously change to the random network cases which power-law is more weak, while small world network of 2-dimemional lattice nearly remains the same curve.  We still need more work here to get a conclusion.

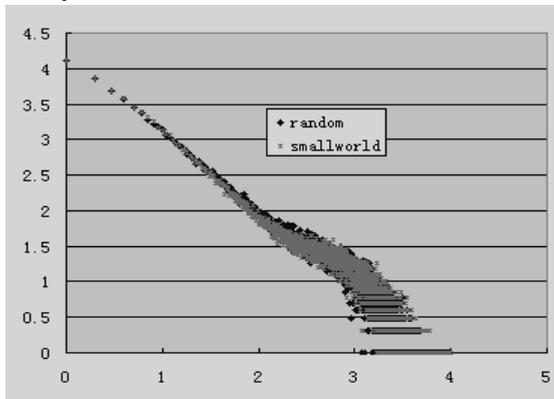
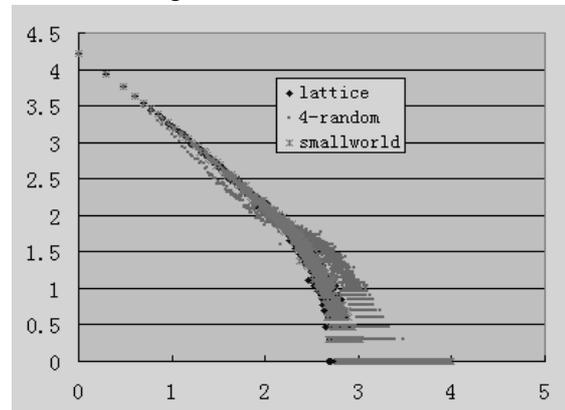

Fig.15 Small world of *hive* compared *random*.       Fig.16  Small world of lattice compared to *random*.
n=100, $f_0$ =0.40, 100000 avalanches                n=100, $f_0$ =0.30, 100000 avalanches

## IV. Preliminary Conclusions and Discussions

In the above section, we have tried to investigate how *system size*, *density of connectivity* and *types of connectivity* affect the power-law behaviors.  It is evident from the simulations that system size, density of connectivity, and types of connectivity can affect the forming of power-law behaviors.  The preliminary experiments show that following facts will affect the power-law behaviors:

1. *System size* (circle structure system):
    a) For all $n \leqslant 3$, it obeys exponential distribution. This can be proved mathematically.
    b) For middle system, the distribution is combined the features of the exponential distribution and those of power-law distribution.
    c) For large size system, power-law is obvious.
2. *Density of connectivity*: higher density, harder to find power-law distribution, which more region is dominated by exponential distribution.
3. *Type of connectivity*:
    a) We can observe power-law behaviors in *hive*, *lattice* network;
    b) But we can't observe power-law behaviors in the *random* network;
    c) For Small World network of those structures that we can observe power-law behaviors, some are like power-law (hive), some are not (2-dimemsional lattice).

We can see that both decrease in system size, increase in connectivity density, and higher clustered (maybe) connectivity types can result in the decrease in locality of the interaction, which will drive the system away from a state with power-law distribution. Systems with long-range interaction may exhibit power-law behaviors in some special circumstances, but the behaviors are not what we care about. The most important feature of complex systems is that local interaction can lead to interesting dynamics.

There is still a lot of future work. We need to perform more simulations on the evolution model to test the liability of, the conjecture, or the tendency we have made in the previous parts of this article. We can also use this kind of approach to study other complex systems, which has different structure of interactions, not limited to BS model. We believe this approach will be very attractive.

**Acknowledgement**
The authors would like to thank for Santa Fe Institute and Central European University for providing a great opportunity to attend the Summer School.